# Synthetic aperture interference light (SAIL) microscopy for high-throughput label-free imaging

*Chenfei Hu[1,2], Mikhail E. Kandel[1,2], Young Jae Lee[2,3], Gabriel Popescu[1,2,*]*

[1]Department of Electrical and Computer Engineering, [2]Beckman Institute for Advanced Science and Technology, [3]Neuroscience Program, University of Illinois at Urbana-Champaign, Urbana, Illinois 61801, USA

[*]Corresponding author: Gabriel Popescu (gpopescu@illinois.edu)

**Abstract:**

Quantitative phase imaging (QPI) is a valuable label-free modality that has gained significant interest due to its wide potentials, from basic biology to clinical applications. Most existing QPI systems measure microscopic objects via interferometry or nonlinear iterative phase reconstructions from intensity measurements. However, all imaging systems compromise spatial resolution for field of view and vice versa, i.e., suffer from a limited space bandwidth product. Current solutions to this problem involve computational phase retrieval algorithms, which are time consuming and often suffer from convergence problems. In this article, we presented synthetic aperture interference light (SAIL) microscopy as a novel modality for high-resolution, wide field of view QPI. The proposed approach employs low-coherence interferometry to directly measure the optical phase delay under different illumination angles and produces large space-bandwidth product (SBP) label-free imaging. We validate the performance of SAIL on standard samples and illustrate the biomedical applications on various specimens: pathology slides, entire insects, and dynamic live cells in large cultures. The reconstructed images have a synthetic numeric aperture of 0.45, and a field of view of 2.6 × 2.6 mm$^2$. Due to its direct measurement of the phase information, SAIL microscopy does not require long computational time, eliminates data redundancy, and always converges.

**Introduction**

Quantitative phase imaging (QPI) is a label-free modality that has emerged as a powerful label-free imaging approach [1]. QPI utilizes the optical phase delay across an object as the contrast mechanism, and thus, it reveals structures of transparent specimens with remarkable sensitivity. Because this modality does not require exogenous labeling, QPI is particularly suitable for non-destructive investigations of biological samples [2]. Optical phase delay is linearly proportional to the non-aqueous content of cells and, thus, it can inform on the cell dry mass, growth, and transport [3-8]. QPI directly reports on the intrinsic biophysical properties of the sample of interest, without artifacts associated with contrast agents. In the past, we have witness a broad range of QPI applications in cellular biology [9-11], cancer pathology [12-14], tissue disorders [15-17], and neuroscience [18-20], as well as materials science [21-24]. More recently, in combination with computational imaging algorithms, QPI enabled tomographic investigations of thin and thick structures [25-29], while deep learning tools enabled new avenues of retrieving chemical specificity from label-free data [30-33].

The throughput of a conventional QPI system is constrained by the space-bandwidth product (SBP), defined by the range of spatial frequencies within a certain spatial volume [34, 35]. Because of this fundamental limit, it is challenging to image a large area with high optical resolution. Although it is generally accepted that the resolution can be improved by using oblique illumination, the factor of enhancement is still limited by the objective numerical aperture, as shown in the *Supplemental Information Section 1* and Ref [36]. Most often, a microscope equipped with a scanning stage is used to accommodate the issue. However, while these instruments are widely spread especially in digital pathology, they require in-focus imaging over large fields of view and some post-processing efforts to align and fuse the individual frames [37, 38]. Fourier ptychographic microscopy (FPM) is an alternative solution [39]. Instead of fusing the images in the spatial domain, FPM boosts the SBP by scanning and fusing portions of the spatial frequency spectrum. Typically, FPM employs an LED array as the light source, and it records a sequence of intensity images under different illumination angles. Using a low-power objective to ensure a large field of view (FOV), high-SBP images are produced through computational reconstruction. However, due to the lack of direct phase information, the FPM reconstruction requires an intermediate "phase retrieval" step, which can be time consuming [39]. Furthermore, fusing the spatial frequency components requires a certain amount of measurement redundancy, *i.e.*, frequency overlap, and nonlinear iterative optimization to retrieve convergent results [40]. A typical FPM uses tens to several hundred input images [41, 42], and the quality of reconstruction is subject to the pattern of illumination [43, 44], sequence of updating [45], LED misalignment [46, 47], and system noise [48].

In this article, we present synthetic aperture interference light (SAIL) microscopy as a novel modality for high-SBP, label-free imaging of micro- and mesoscopic objects. Our imaging system *directly* measures the optical pathlength map associated with the image field under different angles of illumination, which removes the need for computational phase retrieval algorithms. We applied the synthetic aperture (SA) [49-51] algorithm to generate *deterministic* high-SBP QPI images without iterative optimization. Our reconstructed images have a FOV of 2.6 × 2.6 mm$^2$ with a synthetic resolution of 1.45 μm, which is approximately 3× better comparing to the results without SA reconstruction. In contrast to FPM, SAIL reduces the complexity of the reconstruction and data redundancy, while the results can be computed by a CPU processor in a matter of seconds. After the validation on standard samples, we performed measurements on histopathology slides, whole insects, and adherent cell cultures, and compared the results with the ones measured by a conventional QPI microscope. This comparison illustrates how hidden structures in conventional imaging can be easily identified in the SAIL reconstruction. We expect this easy-to-use, high-throughput imaging approach would be of great benefit for biomedical applications in the future.

**Results**

**Principles of SAIL Microscopy**

The general resolution formula for a partially coherent illumination, bright-field microscope, characterized by a condenser numerical aperture, $NA_c$, and objective numerical aperture, $NA_o$, is given by $R = 1.22\lambda / (NA_o + NA_c)$. This classical expression may suggest that one can use a low $NA_o$ to achieve large fields of view, and a high $NA_c$ for high resolution, which would lead to arbitrarily high space-bandwidth products. However, as we show in the Supplemental Information leading to the key Eq. S16, this resolution formula only applies when $NA_c \leq NA_o$, meaning that the traditional expression for resolution with incoherent illumination should be amended to

$$R = 1.22 \frac{\lambda}{NA_o + NA_c} \bigg|_{NA_c \leq NA_o}. \qquad [1]$$

Physically, this expression is due to the image formation being the result of the interference between the scattered and incident light. For incoherent illumination, each scattered field only interferes with its own incident field. The incident light acts as a reference of an interferometer, thus, amplifying the weak scattering signals (since the camera detects their product). For dark field illumination, the incident field is absent, this amplification mechanism vanishes, and the corresponding scattering fields become negligible compared to those produced by bright field illumination. It is this basic phenomenon that motivated the development of synthetic aperture methods, where the scattered light generated by each dark field illumination plane wave is measured separately and combined numerically. It is important to note that when a plane wave fills the condenser aperture, leading to the light being focused into a spot at the sample, all the plane wave components interfere with each other. As a result, even the dark field illuminated scattered light is amplified by the incident light. Of course, this is the case of the laser scanning microscopy, when the resolution is governed by the condenser aperture (typically, another objective), independent of the objective aperture.

In this spirit, we developed the SAIL microscopy instrument to measure the complex image field under various directions of illumination. The optical setup is illustrated in Figure 1a. The system is built as an upgrade to an existing differential interference contrast (DIC) microscope, by using a light engine in the illumination path and an interferometric module at the output port. We replaced the conventional halogen lamp with a programmable light engine, consisting of a commercial display projector and a collecting lens group. To precisely control the angle of illumination, we adjusted the position of the light engine such that the projected pattern is focused at the condenser aperture plane. The inset in Fig. 1a shows a testing illumination pattern and corresponding image measured at the back focal plane of the microscope objective. At the microscope condenser aperture plane, a Wollaston prism is used to generate two orthogonally polarized illumination beams separated by a small angle, which converts into a small transverse shift at the sample plane. Both beams are incident on the sample and generate scattered light. Using a low magnification microscope objective lens, only a small angular spread of the scattered light is collected to form the image field. To measure the phase delay between the two sheared beams quantitatively, we removed the analyzer in the conventional DIC microscope and attached a gradient light interference microscope (GLIM) module to the output port. The design principles of GLIM have been discussed in detail in previous publications [52, 53]. Briefly, a liquid crystal variable retarder (LCVR) modulates the phase shift between the two image fields and a subsequent linear polarizer renders the polarizations of the two fields parallel. As a result, for a given illumination plane wave, of transverse wavevector $\mathbf{k}_{i\perp}$, the interferogram measured by the detector can be expressed in terms of the irradiance, namely (see *Supplemental Information Section 2*),

$$I(\mathbf{r};\mathbf{k}_{i\perp},\varepsilon_n) = |U(\mathbf{r})|^2 + |U(\mathbf{r}+\delta\mathbf{r})|^2 + 2|U(\mathbf{r})U(\mathbf{r}+\delta\mathbf{r})|\cos\left[\Delta\varphi(\mathbf{r};\mathbf{k}_{i\perp}) + \xi(\mathbf{k}_{i\perp}) + \varepsilon_n\right], \quad [2]$$

where $\mathbf{r}$ is the spatial coordinates; $\mathbf{k}_{i\perp}$ is the transverse wavevector of plane wave illumination; $\varepsilon_n = n\pi/2$ is the phase delay introduced by LCVR; $U(\mathbf{r})$ and $U(\mathbf{r}+\delta\mathbf{r})$ are the image field and its sheared replica, respectively, with $\delta\mathbf{r}$ the shear distance; $\Delta\varphi(\mathbf{r},\mathbf{k}_{i\perp}) = \varphi(\mathbf{r}+\delta\mathbf{r},\mathbf{k}_{i\perp}) - \varphi(\mathbf{r},\mathbf{k}_{i\perp})$ is the phase gradient of interest, and $\xi(\mathbf{k}_{i\perp})$ is the constant phase offset associated with the illumination angle $\mathbf{k}_{i\perp}$. By recording 4 intensity images, with $n = 0, 1, 2, 3$, a GLIM image can be extracted using phase shifting interferometry [1, 54], followed by an integration along the sheared direction [55]. Detailed discussions about the system calibration and image reconstruction procedure are included in the *Supplemental Section 3-4*. Figures 1b-c show the intensity images and the regular GLIM image of 3 μm polystyrene microspheres. The broadened line profile in Fig. 1d indicates the system cannot fully resolve the microspheres. The diffraction-limited resolution of the current setup is approximately 4.27 μm. By alternating the direction of illumination, high resolution features can be collected SAIL. Detailed description of the system design and data acquisition can be found in *Methods*.

**SAIL Image Reconstruction**

Consider a static weakly scattering object, the propagation of the optical field, *U*, inside the medium satisfies the inhomogeneous Helmholtz equation,

$$\nabla^2 U(\mathbf{r}) + \beta^2 U(\mathbf{r}) = -\beta_0^2 \chi(\mathbf{r}) U(\mathbf{r}),  \quad [3]$$

where $\beta_0 = \omega/c$ is the wavenumber in vacuum; $\beta = n_0 \beta_0$, with $n_0$ the background refractive index; $\chi$ is the object's scattering potential, which essentially provides the image contrast. *U* is the total field, containing both the incident and scattered light. For a single plane wave incident, the forward scattering field, $U_s$, collected by an imaging system can be derived from Eq. 2 [56, 57], and the solution takes the form (see *Supplemental Information - Section 1* for a derivation)

$$U_s(\mathbf{k}_\perp, z=0) = A(\mathbf{k}_{i\perp}) H(\mathbf{k}_\perp) \chi(\mathbf{k}_\perp - \mathbf{k}_{i\perp}, 0),  \quad [4]$$

where $\mathbf{k}_\perp$ is the transverse spatial-frequencies, *A* is the aperture function of incident light, *H* is the system transfer function. We only consider the object in the focal plane, z=0, and neglect the dependence in the axial direction. Assuming an ideal optical system free of aberration, *H* can be approximated as a 2D uniform disk with the radius of $\beta$NA. Equation 3 states that an illumination field shifts the object's spectrum, and only the frequency components within the system pupil plane are captured to form an image. By employing low-coherence interferometry to capture portions of scattering potential spectrum, synthetic, high-SBP images beyond the conventional limit can be obtained through post-reconstruction.

The principle of SAIL reconstruction is to retrieve synthetic scattering potential, $\hat{\chi}(\mathbf{k}_\perp)$, by stitching the image spectra measured across all illumination angles. This method, generally referred to as synthetic aperture (SA) [49-51, 58], is depicted in Fig. 2. Figure 2a shows a mixture of 1 μm and 3 μm polystyrene microspheres measured at the central illumination, $\mathbf{k}_{i\perp} = 0$. Each SAIL map corresponding to an off-axis illumination angle directly records different portions of the sample's scattering potential. Appending these spectral regions to the regular spectrum ($\mathbf{k}_{i\perp} = 0$), we obtain a synthetic object spectrum (Fig. 2c). By taking a 2D inverse Fourier transform, a SAIL high-SBP image is obtained and its phase extracted, as shown in Fig. 2d. A detailed description of the reconstruction can be found in *Methods*. For the current experimental setup, we achieved a synthetic NA of 0.45, equivalent to 3× enhancement compared to the objective NA. Figure 2e compares the cross-section of the microspheres measured with and without SA reconstruction. Although the synthetized resolution limit cannot resolve features below 1.45 μm, SAIL significantly enhances the sensitivity to the objects below the resolution limit. Compared to existing methods, such as FPM, we measure the complex fields directly, such that the SAIL reconstruction does not require nonlinear iterative optimization to seek convergence, and there is no need to split the FOV into small patches whose areas are on the order of the coherence area [59, 60].

**SAIL microscopy of histology slides**

SAIL directly allows screening of histological specimens with high throughput. To illustrate this capability, we performed measurements on unstained histological specimens. A large FOV image of a breast tissue microarray (TMA) and prostate tissue biopsy are shown in Figs. 3a and 3h, respectively. Using a low-magnification objective, the imaging system directly measures a FOV of 2.6 × 2.6 mm$^2$. Figures 3b-d, and Figures 3i-k show regular inline illumination (GLIM) measurements of tissue portions, while the corresponding SAIL reconstruction at the same sites are displayed in Figs. 3e-g, and Fig. 3l-n, respectively. The direct comparison in Fig. 3 shows that SAIL enhances the resolution, where the shape of the tissue gland and fibers in the extra-cellular matrix can be easily identified after the SAIL reconstruction.

**Imaging of entire organisms**

SAIL is applicable to high-resolution imaging of mesoscopic organisms without scanning. We tested this performance on several fixed insects. Figure 4a shows the wide field image of a female drosophila fixed on a microscope slide. Figures 4b-d show regular inline illumination images zoomed in at the costal margin, basal cell, and post-vertical bristles, respectively. The corresponding SAIL reconstruction zoomed in at the same regions are displayed in Figs. 4e-f. As indicated by the arrows, the computed high-resolution images show the details that cannot be resolved in regular GLIM. In *Supplemental Figure. S4*, we include a similar comparison, where the experiment was performed on a fixed peach worm.

**Measurements on cells *in vitro***

Compared to standard FPM, SAIL microscopy avoids measurement redundancy and provides fast, deterministic reconstruction, with no convergence concerns. Thus, SAIL can be applied to study time-resolved behavior of large cell populations. To illustrate this potential, live Hela cell cultures were prepared and measured by SAIL microscopy with and without the synthetic aperture. The cells were maintained at room temperature (~ 21 °C), such that the abnormal incubation condition would inhibit the function of cellular proteins and eventually trigger necrosis [30]. We took one SAIL measurement every 5 minutes, and continuously monitored the cell culture for 3 hours. Figure 5a show the SAIL phase image measured at $t = 0$. In Fig. 5b. we show the inline illumination image of the cropped region shown by the green rectangle in Fig. 5a (150 × 150 µm² area) measured at $t = 0$, 1.5 hours, and 3 hours, as indicated. The corresponding SAIL images are shown in Fig. 5c. This time-lapse recording shows the cell volume continuously decreasing, indicating occurrence of necrosis. Also, the SAIL images reveal structures in the cell bodies, while the cytoplasm appears to be a smooth surface in the regular inline illumination images. A video of these time-lapse measurements can be found in Supplemental Movie 1. In *Supplemental Figure. S5*, we show a similar comparison, where the measurements were performed on fixed neuronal culture *in vitro*. Thus, the proposed SAIL microscopy allows large population live cell screening with high throughput and sensitivity.

**Summary and Discussion**

We demonstrated SAIL microscopy as a novel approach for achieving high SBP imaging on mesoscopic transparent specimens. The optical system upgrades an existing QPI microscope by employing a light engine to record the quantitative phase delay under different angles of illumination. By applying the SA reconstruction to synthesize image information, we produce high-resolution QPI results while maintaining a large FOV. Our experimental validation shows that SAIL microscopy enables high-resolution imaging of histological slides, mesoscopic organisms, and cells *in vitro* without scanning or iterative nonlinear reconstruction.

Compared to existing methods which generate high-SBP QPI results by iterative reconstruction from intensity measurements, SAIL produces results with rapid phase retrieval and low redundancy. Because the interferometric module directly measures the phase delay across the FOV, the reconstruction procedure avoids the step of iterative phase retrieval, and avoids splitting the FOV into small patches. Furthermore, as each QPI map is an independent measurement of the object's Fourier spectrum, the SAIL reconstruction does not require significant spectrum overlapping between adjacent measurements.

The current SA reconstruction scheme assumes an ideal optical system free of aberration. By adapting aberration estimation algorithms [61-63] into the reconstruction framework, aberration-corrected, high quality images could be retrieved, at the expense of longer computational time. Potentially, more efficient, real-time reconstruction would be possible by transferring the processing onto the video card or applying a deep neural network. On the other hand, the complexity of the microscope alignment could be further simplified by replacing the light engine with a liquid crystal module inserted at the condenser aperture plane. In summary, we expect that this novel label-free, high-throughput imaging modality would be widely used for biomedical research and applications, whenever large areas are studied with high resolution.

**Methods**

**SAIL microscopy configuration and operation**

We built the SAIL instrument to record the quantitative phase and amplitude of the interfering fields under various plane wave illumination. A system schematic is shown in Fig. 1a. The system consists of an existing DIC microscope, a light engine coupled to the illumination port, and an interferometric module attached to the detection port. To precisely control the illumination angle, broadband white light emitted from a display projector (Home Cinema 5030UB, Epson) is first collected by a doublet ($f \approx 285$ mm, $\emptyset = 50.8$ mm, Thorlabs), followed by a second doublet lens ($f = 200$, $\emptyset = 75$ mm, Thorlabs) to form a de-magnified image focused on the microscope pupil plane. To reduce the chromatic aberrations, a bandpass filter (200456, Chroma) is inserted in the optical path, and the resulting light beam has a central wavelength at 535 nm with a bandwidth of 40 nm. We used a commercial DIC microscope (Axiovert 200M, Zeiss) with a 5×/ 0.15 NA objective lens (N-Achroplan 420931-9911, Zeiss), and the Rayleigh resolution is calculated to be 4.35 μm. At the microscope output port, the image fields first entered a LCVR (LCC1423-A, Thorlabs) mounted at a 45° with respect to the input polarizer, followed by an analyzer that renders the polarizations of the image fields parallel. The resulting interferograms are recorded by a sCMOS camera (Zyla 4.2, Andor) with a pixel size of 6.5 μm. The FOV at the specimen plane is approximately $2.6 \times 2.6$ mm$^2$. The procedure of LCVR calibration is included in *Supplemental Note 2*.

By projecting a small circular disk on the condenser aperture plane, we approximated a plane wave illumination. For each SAIL image we selected measurements from 9 plane waves to perform the SA reconstruction. These illumination angles are equally spaced, with the illumination NA of 0.27. The procedure of illumination angle calibration is described in *Supplemental Note 3*. The camera exposure and LCVR stability time were set to be 120 ms and 80 ms, respectively, where the long exposure time is required to measure weak scattering signals generated from oblique illumination angles.

**SAIL reconstruction**

In this project, we consider an ideal optical system free of aberration, and the amplitude of incident field is constant across all illumination angles. Figure 2 summaries the procedure of the reconstruction. First, we up-sample the image under the on-axis illumination ($k_{i\perp} = 0$) as the initial guess for the SAIL output (Fig. 2a). Because SAIL employs common-path interferometry, where the two interference beams travel in the same direction, the resulting phase measurement does not contain the phase ramp associated with the illumination direction [50, 64]. To reverse this process, for each measurement at an oblique illumination angle, the complex image field is multiplied by $\exp(-i\mathbf{k}_{i\perp} \cdot \mathbf{r}_\perp)$ to shift the image spectrum with respect to the illumination direction. Next, we update the output SAIL spectrum by replacing the sub-region with the shifted image spectrum (Fig. 2b). We repeat the previous step until all the off-axis images are updated. For successive images with overlapping regions in the spectrum domain, the average of the overlapping pixels is calculated to update the output spectrum. Finally, by taking a 2D inverse Fourier transform, a SAIL image is obtained by extracting the phase from the resulting image, as shown in Fig. 2c. In this project, a high-resolution SAIL image is computed in MATLAB by a regular desktop with an Intel i7 CPU with a 64 GB RAM. To avoid buffer overflow, we split the wide FOV into 3×3 patches, and the computation time for each sub-FOV takes approximately 6 seconds.

## Data Availability

The data that support the findings of this study is available from the corresponding author upon reasonable request.

## Code Availability

SA reconstruction script with an example dataset is available for download via [https://uillinoisedu-my.sharepoint.com/:u:/g/personal/chenfei3_illinois_edu/Ec1SZc8mkuRGqLaFl0qFH1QBSNXECHih4R4lg5gGRAhhDA?e=UM0O8y](https://uillinoisedu-my.sharepoint.com/:u:/g/personal/chenfei3_illinois_edu/Ec1SZc8mkuRGqLaFl0qFH1QBSNXECHih4R4lg5gGRAhhDA?e=UM0O8y) .

## Acknowledgement

This work is sponsored in part the National Institute of Health (R01CA238191, R01GM129709). C. H is supported by National Institute of Health (T32EB019944).

## Author Contributions

C. H. built the optical system, developed the reconstruction algorithm, collected the data, performed reconstruction and analysis, prepared the figures. M. E. K developed the control software. Y. L. prepared HeLa and neuronal cell culture. C. H. and G. P. wrote the manuscript and GP supervised the project.

**Competing financial interest**

G.P. has financial interest in Phi Optics, a company developing quantitative phase imaging technology for materials and life science applications, which, however, did not sponsor the research. The authors disclosed this invention to the Office of Technology management at the University of Illinois at Urbana-Champaign.

**List of supplemental materials**

1. Supplemental Information
2. Supplemental Movie 1: Time-lapse SA-GLIM and GLIM measurement of HeLa cells *in vitro*.

Figure 1. **SAIL system setup and data collection. a.** SAIL microscopy is an upgrade of an existing DIC microscope with a programmable light engine to modulate the angle of illumination, and a phase imaging (GLIM) module at the detection port. The light engine is properly aligned, such that the projected pattern is focused on the condenser aperture plane. **b.** For each illumination angle, 4 intensity images with different phase delay are recorded to retrieve a QPI map. **c.** low-resolution GLIM measurements of 3 μm polystyrene spheres. **d.** Due to the resolution limit, the cross-section of a bead selected in **c** appears to a broadened profile.

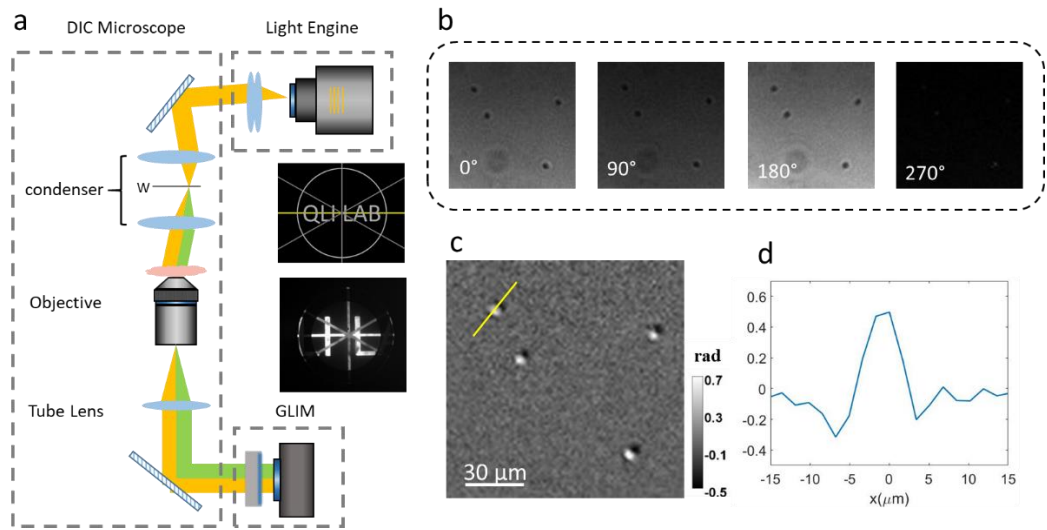

Figure 2. **Principle of the SAIL reconstruction. a.** Using a mixture of 3 μm and 1 μm polystyrene spheres as a testing object, SAIL reconstruction starts with the phase map corresponding to the central illumination. **b.** Taking a 2D Fourier transform of the complex field in **a**, the image is brought to the spatial frequency domain. The dashed circle indicates the location of the objective NA. **c.** For each measurement with oblique illumination angle, the complex field is also Fourier transformed, and the resulting spectrum is appended to the spectrum in **b.** to retrieve a synthetized broad spectrum. **d.** Taking an inverse 2D Fourier transform to the synthetized spectrum, a SAIL reconstruction is obtained by extracting the phase of the resulting field. **e.** Line profiles of the selected microspheres in **d.** before and after SAIL reconstruction.

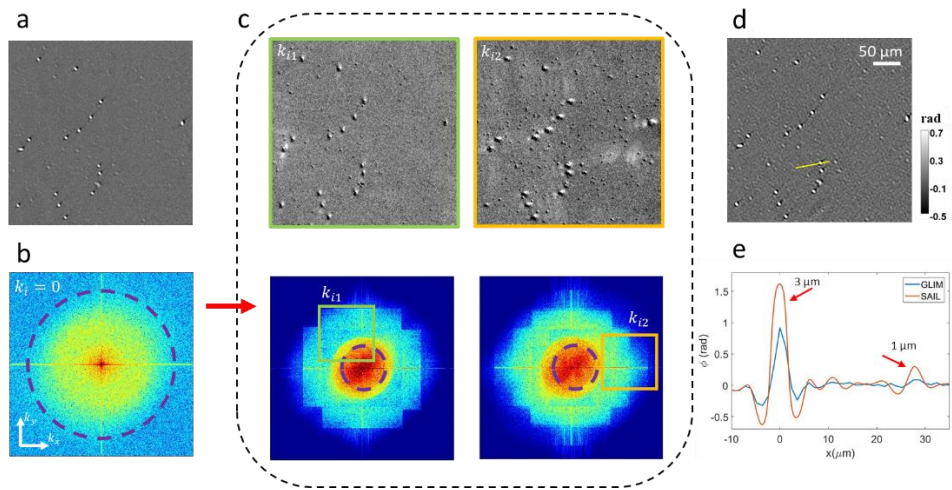

Figure 3. **SAIL imaging of histological specimen. a.** Wide-field low-resolution measurement of a breast tissue microarray, with an imaging area of 2.6×2.6 mm². **b-d.** sub-regions (200 × 200 μm²) cropped from **a**. **e-g.** SAIL reconstruction cropped at the same region as those in **b-d**. **h.** Wide-field low-resolution image of prostate tissue biopsy. **i-k.** sub-regions cropped from **h**. **l-n.** SAIL reconstruction cropped at the same region as those in **i-k**. As indicated by the yellow arrows, SAIL microscopy reveal tissue structures that cannot be resolved by conventional low-resolution QPI.

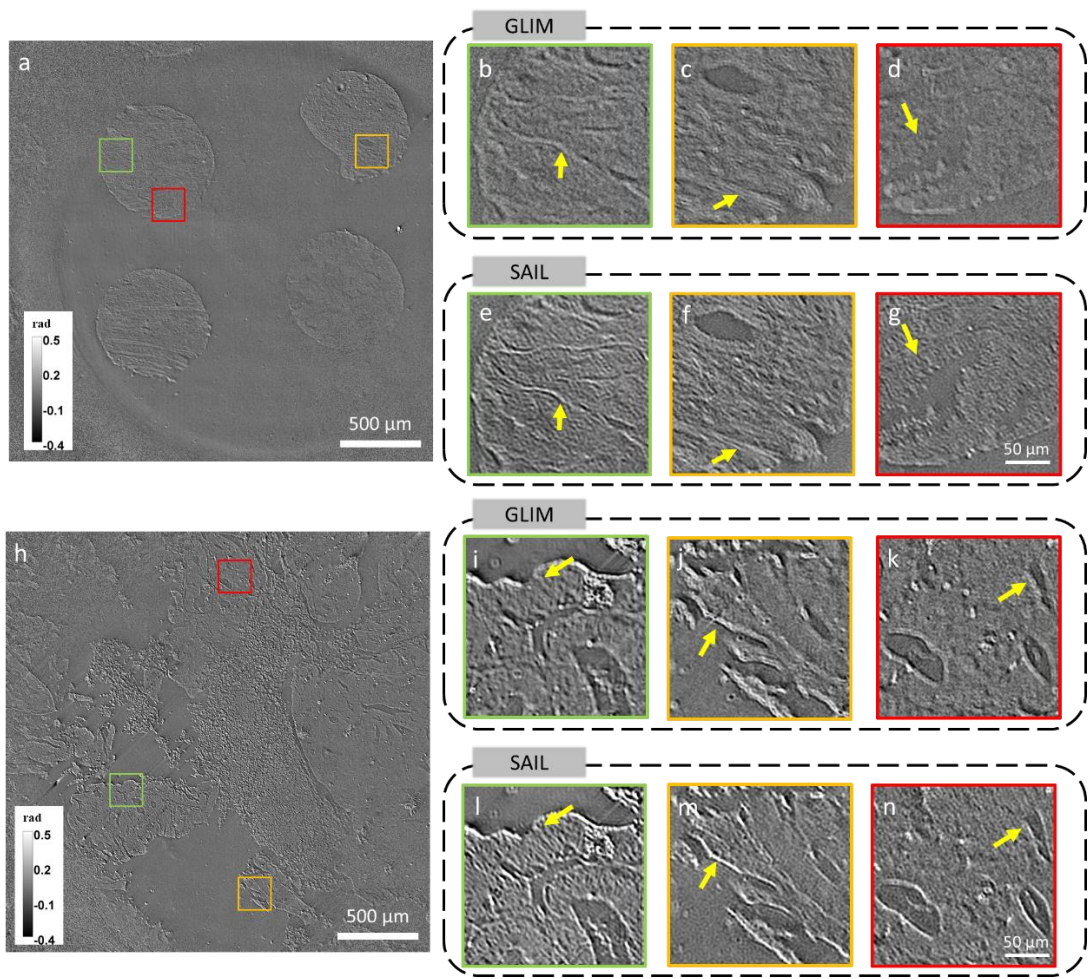

Figure 4. **SAIL imaging of a mesoscopic insect. a.** A wide field image of a female drosophila. **b-d.** zoom-in views of costal margin, basal cell and post-vertical bristles cropped from **a,** respectively. **e-g.** SAIL reconstruction cropped at the same region as those in **b-d**. The yellow arrows indicate more features of costal margin, basal cells and post-vertical bristles can be identified though SAIL reconstruction.

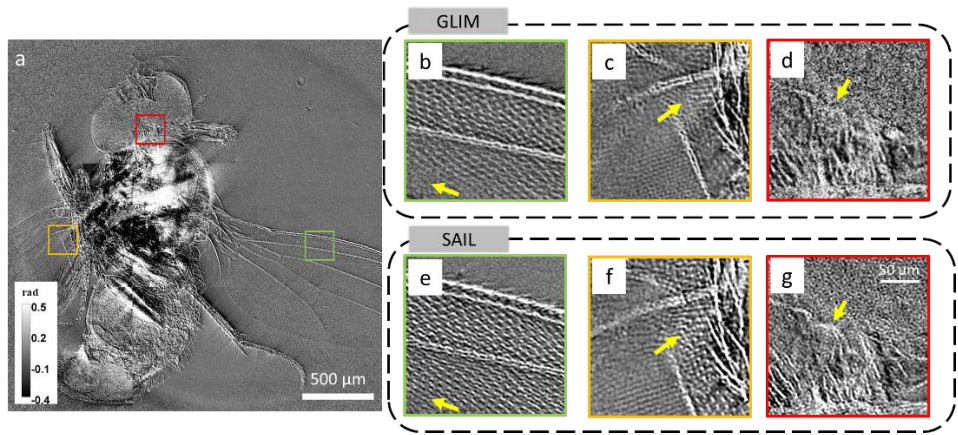

Figure 5. **SAIL imaging of adherent live cells *in vitro*. a.** A wide field measurement of HeLa cells. **b.** Time-lapse GLIM measurements of a sub-region located at the green rectangle in **a,** respectively. **c.** Time-lapse SAIL reconstruction cropped at the same region as those in **b**. A time-lapse video can be found in Supplemental Video 1. The yellow arrows one HeLa appear to be a smooth surface at each time point measured by GLIM, while SAIL shows more features of the cell body.

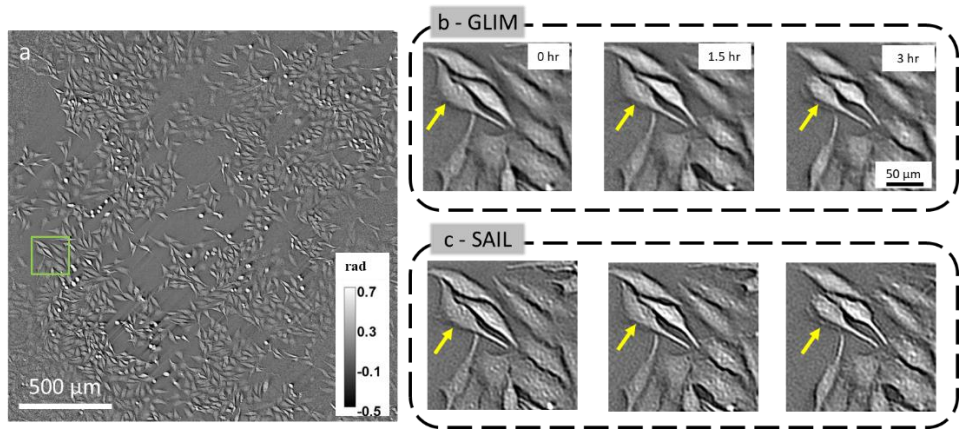